\documentstyle{mn}

\title[Near-Infrared Luminosity Function of Dwarf Galaxies]
{Near-Infrared Luminosity Function and Colours
of Dwarf Galaxies in the Coma Cluster}

\author[B. Mobasher \& N. Trentham]
{Bahram Mobasher$^1$ \& Neil Trentham$^2$\\
$^1$Astrophysics Group, Blackett Laboratory, Imperial College, Prince Consort Rd
, London SW7 2BZ, UK\\
$^2$ Institute for Astronomy, University of Hawaii, 2680 Woodlawn Drive, 
Honolulu HI 96822, U.S.A.\\
}

\begin{document}
 
\maketitle
\begin{abstract}
We present $K$-band observations of the low-luminosity galaxies
in the Coma cluster, which are responsible for the steep upturn in
the optical luminosity function at $M_R \sim -16$, discovered recently. 
The main results of this study are
\vskip 1pt
\noindent (i) The optical$-$near-infrared
colours of these galaxies imply that they are dwarf spheroidal galaxies.
The median $B-K$ colour for galaxies with
$-19.3 < M_K < -16.3$ is  
3.6 mag. 
\vskip 1pt
\noindent (ii) The $K$-band luminosity function in the Coma cluster is not well
constrained, because of the uncertainties due to the
field-to-field variance of the background.
However, within the estimated large errors,
this is consistent with the $R$-band
luminosity function, shifted by $\sim3$ magnitudes.
\vskip 1pt
\noindent (iii) 
Many of the cluster dwarfs lie in a region of the $B-K$ vs.~$B-R$
colour-colour diagram where background galaxies are rare
($B-K < 5$; $1.2 < B-R < 1.6$).
Local dwarf spheroidal galaxies lie in this region too.
This suggests that a better measurement of the $K$-band
cluster luminosity function can be made if the field-to-field
variance of the background
can be measured as a function of colour, even if
it is large.  
\vskip 1pt
\noindent (iv)
If we assume that none of the galaxies
in the region of the $B-K$ vs.~$B-R$ plane given in (iii) in
our cluster fields are background,
and that all the cluster galaxies with $15.5 < K < 18.5$ lie
in this region of the plane, then we measure 
$\alpha = -1.41_{-0.37}^{+0.34}$
for $-19.3 < M_K < -16.3$, where $\alpha$ is the logarithmic slope
of the luminosity function. 
The uncertainties in this number come from counting statistics.
\end{abstract}


\section{Introduction} 
Recent studies of the optical luminosity function (LF) of galaxies in the
Coma cluster (Trentham 1997a; 
Bernstein et al.~1995; Secker \& Harris 1996)
have revealed a steep rise at the faint-end, with 
$-1.7 < \alpha < -1.3$ for 
magnitudes 
fainter than about $M_R = -15$ (where $\alpha$ is the logarithmic slope
of the LF: $\phi(L) \propto L^{\alpha}$). 

The optical
colours and scale-lengths of these faint galaxies suggest that they
are probably
dwarf spheroidal (dSph, alternatively called dwarf elliptical)
galaxies.
These galaxies
form a distinct family of objects, separate from giant ellipticals on the
luminosity $-$ surface-brightness $-$ radius parameter correlations
(Kormendy 1987, Binggeli 1994). 
They have lower surface-brightnesses as their luminosity decreases;
well known examples in the Local Group are NGC 205 and Draco. 
The colour distribution of the galaxies which produce a steep rise in
the LF is heavily peaked at $B-R = 1.3$.  This is towards the blue
end of the range of colours exhibited by local dSphs (Trentham 1997b), but
is redder than most of the dwarf irregulars (dIrrs).  
We cannot distinguish between the two types of galaxies based on their
morphologies because (i) they have similar scale-lengths as a function
of luminoisty (Binggeli 1994), and (ii) these scale-lengths are 
comparable to the seeing for galaxies in the Coma, so we cannot probe fine
details in the structure.

Here, we extend these studies to the near-infrared wavelengths. The K-band
measurements probe the old stellar populations in galaxies as opposed to
the younger population probed
by the optical wavebands. 
The aim of this study is two fold: 1). 
to measure the near-infared LF of galaxies in Coma 
and ascertain whether or not the steep rise found at
optical wavelengths is seen in the $K$-band, and  
2). to study the nature of galaxies which dominate
the faint-end of the LF (ie. dwarfs), using their optical-infrared colours. 

The measurement of the
near-infrared LF is not an easy task because
of background contamination.
At optical wavelengths, the background counts have been well
characterized because CCDs cover substantial areas (e.g.~Bernstein
et al.~1995).  In the near-infrared wavelengths, due to
the smaller format of the arrays, 
such detailed measurements cannot be made.  A number
of medium-deep and deep surveys (Gardner et al.~1993) have permitted
detailed measurements of the mean number counts.
However, these measurements do not constrain the distribution of the
counts around this mean, from field to field as a function of
angular size.
We compute this
distribution from a combination of optical and $K$-band
observations,
and describe this calculation in detail.
This is an important part of this study because the field-to-field
variance of the background
is the dominant source of uncertainty in the LF.

The positions of the cluster dwarfs on
an optical-infrared colour-colour diagram allow
us to identify the type of galaxy
responsible for the upturn in the optical LF, with
somewhat more
confidence than using the optical observations
alone (see above).
The difference in optical-near infrared colours
between dSph and dIrr
galaxies is
much bigger
than the difference in optical colours, because the dSphs,
unlike the dIrrs, have a substantial fraction of the old stellar populations.
The similarity in the optical colours between the bluest dSphs
and the reddest dIrrs is
normally attributed to recent star formation in
the blue dSphs; a plausible physical
mechanism for this is given by Silk et al.~(1987).

This paper is organized as follows.  In Section 2, we describe our observing
strategy, the observations and data reduction. 
Section 3 presents the background counts. The near-ir LF for galaxies in the 
Coma cluster is studied in Section 4. The nature of the dwarf galaxies in
the Coma cluster is explored in section 5, using their optical-infrared colours.
Finally, our conclusions are summarised in section 6.

Throughout this paper we assume 
$H_0 = 75$ km s$^{-1}$ Mpc$^{-1}$ and
$\Omega_0 = 1$.

\section{Observations and Data Reduction}

\subsection {The Near-infrared Observations}

The requirement to survey a large area of the Coma cluster to deep
levels is prevented by the size and sensitivity of near-infrared 
detectors.  For this study, we wish to cover enough area to
get good counting statistics after background subtraction,
while going deep enough that we can still 
find the faintest dwarfs.
Our observing strategy reflects these two requirements. 

We carried out a survey in two parts.
A wide-angle shallow ($K \sim 19$) survey of the Coma cluster 
core was 
performed using the $1024\times 1024$ HgCdTe 
detector (QUIRC) at the University of Hawaii 2.2 m Telescope.  
This was then complemented by a deep survey 
covering a much smaller 
area in the core to fainter limits ($K \sim 21$), using the
$256\times 256$ InSb array IRCAM3 at the UKIRT 3.8 m telescope. 
The areas we surveyed are shown in Figure 1. 

We also surveyed an area around the NGC 4839 galaxy at both the UH 2.2 m and
the UKIRT.  This galaxy is at the center
of a small group, 41 arcminutes to the southwest of
the cluster center, 
and is perhaps in the process of merging with the
Coma cluster.  
Details of the observations and photometry are discussed in the following 
sub-sections.

\subsection{The QUIRC (UH 2.2m) Observations and Data Reduction}

The observations here were taken 
at the f/10 Cassegrain focus of the University of Hawaii 2.2 m telescope
on Mauna Kea during the nights of March 11$-13$ 1995. 
The detector was the QUIRC  
1024 $\times$ 1024 HgCdTe array 
(scale 0.19$^{\prime\prime}$ pixel$^{-1}$,
field of view
3.2$^\prime$ $\times$ 3.2$^\prime$).

A total of 41.1 square arcminutes of the cluster core (see
Figure 1) 
and a 9.9 square arcminute region, centered on NGC 4839 
($\alpha (1950) = 12^{\rm h} 54^{\rm m} 59.0^{\rm s}$,
$\delta (1950) = 27^{\circ} 46^{\prime} 3^{\prime \prime}$) 
were imaged, using the $K^{\prime}$ filter (Wainscoat \& Cowie 1992).
The imaging of the core region was done as a 3$\times$2 mosaic of  
individual fields.  Exposure times range from 60 mins to 36 mins for the core 
fields and the NGC 4839 field respectively.  The seeing FWHM 
varied from 0.9$^{\prime\prime}$ to 1.1$^{\prime\prime}$.

Each of the six core field images and the single NGC 4839 image
were constructed from a number of three-minute exposures, taken in 
sets of six, with sky exposures taken between different sets.  
These images, each of three-minute exposures, 
were dark and sky subtracted, flatfielded (dome flats
were used), and then registered and coadded. 
The sky fields were chosen to lie in 
nearby uncrowded regions, and
were dithered and median filtered as sets of six two-minute
exposures to remove objects in the fields and
cosmic rays. 
This procedure produced images flat to better than one percent.

Instrumental magnitudes were then computed from observations of several
($\sim 8$ per night) standard stars 
from the UKIRT Faint Standards list of 
Casali \& Hawarden (1992)
and the photometry was converted to
the $K^{\prime}$ system of Wainscoat \& Cowie (1992), using their relation
$K^{\prime} - K = 0.22 (H-K)$. 
The zero-point is accurate to
about 3\%.  

We also took images of one blank
sky field (9 square arcminutes, 
centered on $\alpha (1950) = 15^{\rm h}$,
$\delta (1950) = 40^{\circ}$) 
for a total integration time of
60 minutes. 
These images were reduced as above, except that the individual
exposures themselves were median filtered to construct a sky
image for subtraction (this was possible because there were no
bright galaxies present).
This field will be used to calculate the background counts in the 
following section, to correct the cluster galaxies for background
contamination.

\subsection{The IRCAM3 (UKIRT) Observations and Data Reduction}

The deep observations here were taken
at the 3.8 m United Kingdom Infrared Telescope (UKIRT)
on Mauna Kea during the nights of March 29$-31$ 1995,
using the IRCAM3 
256 $\times$ 256 InSB array.
(scale 0.286$^{\prime\prime}$ pixel$^{-1}$,
field of view
1.2$^\prime$ $\times$ 1.2$^\prime$).

A total of 1.1 square arcminutes of the cluster core 
was imaged (see Figure 1) using a $K$ filter. 
We chose a region that was devoid of very bright galaxies.
This $K$ filter is slightly different from the $K^{\prime}$ filter we
used for the QUIRC observations; the magnitude zero-point
offsets are given by the equation in the previous section.
We also imaged a similar size region 
centered on NGC 4839.
The mean seeing was 1.0$^{\prime\prime}$.
The total integration times were 6 hours for each field.
These images reach $K = 20.5$ ($M_K = -14.3$ for galaxies
in the Coma cluster assuming a distance modulus of 34.8 mag. for the Coma). 
This provides the deepest near-infrared survey of a 
cluster presently available.
Instrumental magnitudes were again computed
from observations of the
UKIRT Faint Standards.

The observing strategy was similar to that described in
Section 2.1.1.
Each exposure here was 60 seconds, and there were 27 such 
exposures per set, covering the area around the center of each field. The
flatfield frames were constructed using the median filter of the sky frames. 
For the purpose of sky subtraction, a sky field was chosen to 
lay outside the cluster (at about
4$^{\circ}$, corresponding to 30 times the core radius).
All the exposures of this field were combined 
to make a single deep image.
This deep image will also be used to
compute the background number counts 
in the
following sections.  
It is sufficiently well outside the cluster to be
safely used for these purposes.

\subsection{Optical CCD Observations}

The optical measurements of the cluster fields used
in the following sections are  
from the UH 2.2 m $B$ and $R$-band survey of Trentham (1997a).
The details of the observations and reduction are described 
in that paper. 
The observations are based on
the Johnson (UBV) $-$ Cousins (RI)
magnitude system of Landolt (1992).

The optical images of the $15^{\rm h}$
$40^{\circ}$
background fields were taken during service time on the INT 2.5 m 
telescope on La Palma
with a Tektronix 1024 $\times$ 1024 CCD.  
The exposure times were 30 minutes each in $B$ and $R$;
the data reduction was carried out in an identical way to
the UH 2.2 m survey data. 

\subsection{Source Detection and Photometry}

Using the reduced near-infrared frames, 
objects were detected at the 3$\sigma$ level above the
background, using the detection algorithm FOCAS 
(Jarvis \& Tyson 1981; Valdes 1982, 1989). Their isophotal magnitudes
were then estimated and a 
surface-brightness dependent technique was developed to 
correct them to the total magnitudes.
The technique
is described in detail elsewhere (Trentham 1997c)
and here we only give a summary of the main steps:

\noindent {\bf a)} We measure the sky rms noise, $\sigma_{rms}$, and
the FWHM seeing, $b_{\rm FWHM}$, for each image.
We then simulate galaxies of various
apparent magnitudes and exponential scale-lengths,
convolve them
with a gaussian seeing function of width 
$b_{\rm FWHM}$ and add
Poisson noise of rms magnitude $\sigma_{rms}$. 
Next, the FOCAS detection algorithm was run on this image, 
searching for objects larger than
one seeing disc 
with fluxes 3$\sigma_{\rm rms}$ above the sky.
For each object detected
we measure the isophotal magnitude $m_I$ and its
first-moment light radius
$r_{1}$. 
As the true magnitude $m$ of each object is known, we compute
the function $m(m_I, r_1)$, and its uncertainty  
$\sigma (m)[m_I, r_1]$ for each image.
The uncertainty is estimated by investigating how intrinsically
identical galaxies are detected differently, depending on
the local noise.   
We also determine the faintest magnitude $m_L$, at
which galaxies with intrinsic magnitudes and scale-lengths
equal to those of the local dwarfs, projected to the
distance of the Coma, are detected with 100\% completeness
(the magnitude
vs.~scale-length relation for dwarfs was computed using the
data of Kormendy 1987).
This will be the faintest magnitude to which we determine
the LF in each image.  
\vskip 1pt
\noindent {\bf b)} We then run the same detection algorithm
on our data and make a catalog of all the objects detected
at the 3$\sigma$ level (typically 21 $K^{\prime}$ 
mag arcsec$^{-2}$ for the QUIRC images and
23 $K$ mag arcsec$^{-2}$ for the IRCAM3 images) 
above the sky, measuring their $m_I$ 
and $r_{1}$ parameters.  The FOCAS splitting algorithm, which
searches for multiple maxima within a single isophote, is
used to identify and separate the merged objects.
\vskip 1pt
\noindent {\bf c)}
Objects are then classified 
as ``galaxies'' or ``stars'' (see Valdes 1989)
based on their morphology relative to that of
several reference PSF stars in the field.
At faint magnitudes ($K > 18$), these
classifications are unreliable because many galaxies
have apparent scale lengths
smaller than the seeing and so, look like stars.   
Therefore, we correct for stellar contamination at faint
magnitudes by
computing the number of faint stars expected, given the number
of bright ($K < 17$) stars and assuming some Galactic stellar
luminosity function slope (we adopt the slope measured at
optical wavelengths 
by Jones et al.~1991).   
For the Coma, which  lies well out of the Galactic plane, this
is a small effect ($<5$\%) and the uncertainties generated by this
method are negligible.
\vskip 1pt
\noindent {\bf d)} 
>From our measured $m_I$ and $r_1$ values, 
for each object in our catalog, we
then compute $m$ and its uncertainty $\sigma (m)$,
bin them in half-magnitude (QUIRC) or one-magnitude (IRCAM3) intervals,
and correct for stellar contamination as in (c) above.
The number counts (in units of number per magnitude per square degree)
are then computed by dividing the number of 
galaxies in each bin by the area of the survey. 
The uncertainties in the number counts is
the quadrature sum of counting statistics and uncertainties from the
isophotal corrections $\sigma (m)$.
This surveyed area includes a correction for crowding, the
process by which faint galaxies go undetected because they happen to
fall within the detection isophote of a much brighter object.

For each object, we also compute the aperture magnitude
$m_a$ within an aperture diameter of 3.0$^{\prime\prime}$. The colours of 
the faint
objects are measured by taking the difference between their aperture
magnitudes in different bands.
Isophotal magnitudes are not used for computing colours 
because the detection isophotes are not the same in
different bands.
The selected aperture here is large enough that differential seeing effects 
between different images are negligible. 
This method fails to give accurate colours for bright galaxies if
colour gradients are large. However, in this study only
colours for the fainter galaxies ($K > 15.5$) will be explored.

The $K$-band
Galactic extinction is negligible for both the cluster and background
fields here (based on their positions in the HI maps of
Burstein \& Heiles 1982) and hence, has been ignored in this investigation. 
The above technique was applied to both our cluster and background fields. 
The total magnitudes and colours for galaxies detected in both the QUIRC and
IRCAM3 surveys were then estimated. 

\section{The $K$-band Number Counts}

In this study, we
surveyed two background fields, one with QUIRC (the
15$^{\rm h}$ 40$^{\circ}$ field $-$ see Section 2.1.1), and
one with IRCAM3 (the sky field 4$^{\circ}$ outside the Coma
core $-$ see Section 2.1.2). 
In Figure 2, we present the background number counts for
our fields, compared with
the number counts from the literature as compiled by Gardner et al.~(1993).
Here, all the photometry is converted to the $K$ system, to ensure internal
consistency.  Our data agrees well with that of previous studies, but
our errors are larger.  The dashed line represents the
mean number counts, computed using {\it all} the data in Figure 2. 
The number counts corresponding to this line will be used to correct the
counts in the Coma cluster for background contamination.

The uncertainty in using the mean value of $\Delta N(K)$
for a field size probed by our survey 
(i.e.~the field-to-field variance in the background counts) 
must be quantified, as this will be
a major source of uncertainty in the LF.  This has not been measured
directly in the $K$-band, because of the small size of the arrays.
It has, however, been measured at optical wavelengths (see 
Bernstein et al.~1995 and Trentham
1997c for an estimate of $\Delta N(R)$ and $\Delta N(B)$).  
Also, the distribution
of $B-K$ colours for galaxies as a function of $K$ has been measured
(Gardner et al.~1993).  We 
therefore estimate, for fields having
the same area,
$$\Delta N(K) = N(K) {  
{\int_{B} \left({\Delta N}\over{N}\right)_{B} f(B-K){\rm d}B }\over{
\int_{B}f(B-K){\rm d}B} 
},\eqno(1)$$ 
where $f(B-K)$ is the distribution of $B-K$ colours for a given $K$.
We approximate $f(B-K)$ to a gaussian with the mean
value 
given by Gardner et al.~(1993)- (see their Fig. 3), and a standard
deviation of 1 magnitude. The values of 
$\left({\Delta N}\over{N}\right)_{B}$ are taken from 
Trentham (1997c).

We now calculate $\Delta N(K)$ from Equation 1, correcting
for different area sizes, using Poisson statistics 
(Figure 3).  For the
Coma core field, this correction is small, since the total field size of
our mosaic is comparable to the field size used to derive $\Delta N$
at optical wavelengths. 

\section{The Near-infrared Luminosity Function in the Coma Cluster}

The number of galaxies detected in each K magnitude bin in the Coma cluster
are estimated for both the QUIRC (Table 1) and IRCAM3 (Table 2) observations. 
The fields have different magnitude limits because the seeing and
exposure time varied. 
This is the reason why the numbers 
of fields in the last rows of Table 1 (core section) are
smaller than in the other
rows. The K-band number counts for galaxies in the Coma cluster 
are shown in Figure 4, 
along with the mean background
line described in the previous section.  
 
The luminosity function is computed
by subtracting the background contribution from the cluster
number counts. 
The associated uncertainty in the LF takes into account both the
field-to-field variance in the background, as shown in Figure 3, 
and the errors
described in Section 2.3 and shown in Figure 4.

The near-infrared LF for the Coma core field is 
presented in the upper panel of Figure 5 (a distance modulus of 34.8 mag. 
is assumed for the Coma cluster). After subtracting the background 
contribution, some of the magnitude bins are left with no galaxies 
(i.e. negative number in that bin) and
therefore, on this logarithmic plot, only the upper errorbars are shown. 
The most obvious feature of the LF is its poor statistics.  The errors
are the quadrature sums of the uncertainties from counting statistics,
measurements of the total magnitudes, and the field-to-field variance
of the background.
At the faintest magnitudes, uncertainties due to
background contamination dominates (so, changing the binning will not help).  
This problem will remain until 
it is possible to image much greater areas of the
cluster to fainter limits. The implication here is that
a well-constrained 
measurement of the K-band LF for dwarf galaxies in Coma 
is not
feasible at present, at least if this is carried out
by performing
a simple background subtraction. 
Morphological information does not help since
the apparent scale-lengths of the Coma dwarfs are similar to those
of background late-type galaxies, and are close to the seeing disk. 
A more instructive approach  
might be to make a background subtraction, 
taking into account the colours of galaxies.
This 
requires detailed measurement of the field-to-field
variance of the background counts as a function of colour 
(concentrating
on colours similar to those of the cluster galaxies). 
This is not
a trivial measurment, but is 
easier than the wide-field surveys of clusters
to faint magnitudes described above. However, the variance cannot be 
estimated, using a strategy similar to 
that outlined in Section 3, 
because the contribution to  
$\left({\Delta N}\over{N}\right)_{B}$, by galaxies of different colours, 
is not known.

The shape of the $K$-band LF, within the errors, is roughly consistent
with that of the optical ($R$-band, shifted by about 3 magnitudes) 
LF for the Coma core field.  
The IRCAM3 observations, which probe
a magnitude range where the optical LF is steeply increasing,   
are systemtatically higher than the QUIRC points but
this, in part, could also be
a normalization effect (the IRCAM3 and QUIRC
surveys covered different areas).  
The large error bars mean that 
the parameters from any fit to our $K$-band data will be
poorly constrained.
A Schechter (1976) function fit to the QUIRC data, where we fix
$M_K^{*} = -24.35$ (Barger et al.~1996) gives

$$\alpha^{*} = -1.10,\ \log_{10} N^{*} = 2.49,\ 
\nu = 18,\ {{\chi^{2}} \over{\nu} } = 0.58.$$

\noindent
Alternatively, a power-law fit 
for the range $-20.3 < M_K < -16.3$ gives
$\alpha = -0.83_{-0.99}^{+1.83}$. 

For the NGC 4839 field, the problems of large uncertainties due to
background subtraction are worse because of the smaller field size and 
hence, a
lower galaxy density.  The $K$-band LF in this region is essentially 
unconstrained.

Finally, we stress that in spite of the above difficulties, Coma is still
the best cluster for measuring the $K$-band LF of dwarf galaxies.
For more distant, but richer clusters
like Abell 665 ($z=0.2$),
we have many more cluster galaxies
relative to the background and hence, the LF is less sensitive to 
$\Delta (N)$. However, we cannot realistically probe the faintest
magnitudes in these clusters (where the dwarfs are numerous).
For more nearby clusters like the Virgo, we can probe the
faintest absolute magnitudes, but then covering an area large enough that
counting statistics are reasonable is difficult, considering the
field of view of the near-infrared array detectors. 

\section{The Optical-Infrared Colours of Dwarfs in Field and Clusters}

The $B-K$ vs.~$B-R$ colour-colour diagram for the QUIRC data is shown in 
Figure 6. The objects with $B-K > 5$ are expected to be background late-type 
galaxies. 
Most of the galaxies in the 15$^{\rm h}$ 40$^{\circ}$ field, and many of the
galaxies in the Coma core field reside in this part of the diagram. 

The cluster galaxies are most likely the condensation seen in
the figure with $1.2 < B-R < 1.6$,   
based on their optical colours (Trentham 1997a).
These galaxies have $B-K$ colours somewhat redder than that of the dIrr 
galaxies at any
$z$. This confirms our conclusion from the optical observations, that the
faint galaxies in the Coma are dSphs.
A histogram of the $B-K$
colours of galaxies with $1.2 < B-R < 1.6$ is presented in Figure 7.
This histogram extends from $B-K = 3$ to
$B-K = 5$, peaking at $B-K = 3.6$. 
The vertical error bars in this figure are the
$\sqrt{N}$ counting errors and are large enough 
that conclusions about the detailed shape
of the histogram cannot be made.
Nevertheless, the $B-K$ colour distribution 
seems broader than the $B-R$ distribution
(Trentham 1997a).  This observation can be explained if the dSphs in the Coma 
cluster have a more
homogeneous history regarding their most recent star formation burst
(which is probed by the $B-R$ colour) compared to their
integrated star formation history (which is probed by the $B-K$ colour).
Such a scenario is expected if the Coma cluster is a merger between two
smaller clusters (as suggested by its lack of a cooling flow and its two
brightest cluster galaxies) and if star formation was induced in most of
the low mass galaxies at this time (see Silk et al.~1987).  
 
The absence of background galaxies in the part of the
colour-colour diagram where the cluster dwarfs reside is encouraging. 
This confirms our conclusion in Section 4 that using colour information
to make a background subtraction (to measure the $K$-band cluster LF)
may well be feasible.  Such a strategy requires an extension of 
the measurements presented in Figure 6 to a larger magnitude
range to improve the background statistics by observing a number
of background fields in 3 colours.   
For each magnitude bin, a plot identical
to Figure 6 can be constructed and used to perform the background  
subtraction based on a local ``density on the
$B-K$ vs.~$B-R$ plane'' criterion.  The field-to-field variance would
probably still be large, but the fractional contamination of the
background will be lower, so the LF would be much better 
constrained.    
 
This prescription is implemented
and the results presented 
in Figure 8, where the LF for all the galaxies
with $B-K < 5$ and $1.2 < B-R < 1.6$ in Figure 7 is shown. 
There are no galaxies in the 15$^{\rm h}$ 40$^{\circ}$ background
field in this part of the $B-K$ vs.~$B-R$ plane, and we assume zero
background contamination in our cluster fields.
Only counting statistics are used in
deriving the uncertainties.  The estimated LF here is still not well
constrained; a power-law fit to the data  
gives $\alpha = -1.41_{-0.37}^{+0.34}$.  This is close to the
values measured by Secker \& Harris (1996) and Trentham (1997a) for
this part of the LF (assuming $R-K \approx 3$), but the
errors are still large,
and are likely to be underestimates.
The errors must increase 
as we take the background into account. A more accurate estimate of the
uncertainties here needs more wide-field 
optical/near-infrared surveys and hence, a characterisation of the background
variance, as described in Section 4. 
Also, the cluster members that do not lie in this part of
the $B-K$ vs.~$B-R$ plane are missed, introducing a selection
effect.
However, with a few times the area of this survey, the counting
errors become manageable, and if the background contamination really
is small, a measurement of the $K$-band LF seems feasible. 
Such a survey will be possible shortly, when larger 
infrared arrays become available. 

\section{Conclusions} 

A K-band survey of the Coma cluster has been carried out. This 
consists of a wide-angle shallow ($K\sim 19$) survey with the
QUIRC (UH 2.2m) and a deeper survey ($K\sim 21$), covering a 
much smaller area with the IRCAM3 (UKIRT). These observations
were used to construct the near-infrared LF of galaxies in the
Coma cluster and to study the nature of the population dominating
the faint-end of the LF. The results of this study are summarised 
as follows:

\begin{enumerate} 
\item The K-band LF in the Coma cluster is not well constrained. However, 
within the estimated (large) errors, this is consistent with the R-band LF, 
shifted by $\sim 3$ magnitudes. 

\item The optical-infrared colours of these faint galaxies confirm that
they are dwarf spheroidals. The median $B-K$ colour for galaxies
with $-19.3 < M_K < -16.3$ is 3.6 mag.

\item Using the $B-K$ vs. $B-R$ colour-colour diagram, a region on this plane
is identified where the background galaxies are rare ($B-K < 5$; 
$1.2 < B-R < 1.6$).  It is proposed that a more accurate measurement of
the K-band LF for clusters can be carried out if the background contamination
is estimated as a function of colour.

\item Using the $B-K$ and $B-R$ colours to correct the Coma cluster data
for background contamination (as in 3 above), the K-band LF is
again constructed. The logarithmic slope of the LF in the range
$-19.3 < M_K < -16.3$ mag. is $\alpha = -1.41_{-0.37}^{+0.34}$. 
\end{enumerate}

\section{Figure Captions}

\noindent
{\bf Figure 1:} The Palomar Sky Survey image of the core of the Coma cluster.
The image is 11.2 arcminutes square; north is up and east is to the left.
The black (UH 2.2 m) and white (UKIRT) boxes ouline the regions covered by
this survey.
The two bright central galaxies are NGC 4874
($\alpha (1950) = 12^{\rm h} 57^{\rm m} 11.0^{\rm s}$,
 $\delta (1950) = 28^{\circ} 13^{\prime} 46^{\prime \prime}$)
and NGC 4889
($\alpha (1950) = 12^{\rm h} 57^{\rm m} 43.3^{\rm s}$,
 $\delta (1950) = 27^{\circ} 46^{\prime} 3^{\prime \prime}$).

\vskip 10pt
\noindent
{\bf Figure 2:} The magnitude $-$ number count relation for the $K$-band.  The
solid symbols represent the data from this study (one QUIRC field, and one
IRCAM3 field, as described in the text).  The
$K$-band magnitudes are used. The QUIRC data have been corrected to this
system assuming $H-K = 0.5$ for typical field galaxies and 
$K^{\prime} - K = 0.22 (H-K)$.  
The open symbols refer to the counts from a number of studies, taken from 
the compilation of Gardner et al.~(1993, see also Gardner 1992).  
The abbreviations in the key are
HDS = Hawaii Deep Suvey (see also Cowie et al.~1994);  
HMDS = Hawaii Medium Deep Survey; 
HWDS = Hawaii Medium Wide Survey;
GPCM = Glazebrook et al.~1994; 
HWS = Hawaii Wide Survey.   
The dashed line shows the mean value of $N$ in K magnitude bins, 
computed from all the data presented here.

\vskip 10pt
\noindent
{\bf Figure 3:} The field-to-field variance in the background for
the $K$-band, computed as described
in the text.  The kink at $K\sim 17$ corresponds to the
discontinuity in the mean value of $f(B-K)$
(see Figure 3 of Gardner et al.~1993).

\vskip 10pt
\noindent
{\bf Figure 4:} The magnitude $-$ number count relation for our Coma core
(upper panel) and NGC 4839 field (lower panel) data, computed as described
in the text.  The magnitudes here are $K^{\prime}$ for the QUIRC data and
$K$ for the UKIRT data.    
The dashed line represents the mean $K^{\prime}$-band number counts 
as a function of magnitude, computed from the equivalent $K$-band line
in Figure 2, assuming $K^{\prime} - K = 0.22 (H-K)$ for a typical field
galaxy.   

\vskip 10pt
\noindent
{\bf Figure 5:} The $K$-band luminosity function of
the Coma core (upper panel) and NGC 4839 group
(lower panel).  
The $R$-band data is from Trentham (1997a) and has been
normalized to minimize the scatter between $N(K)$ and
$N(R+3)$.  The magnitude system is $K^{\prime}$ for 
the QUIRC data and $K$ for the UKIRT data (last two
points only in the upper panel and last point only in
the lower panel).  We do not correct all the data to
a single magnitude system because of our lack of
knowledge of $H-K$ for the cluster dwarfs. 
Slopes corresponding to two values of $\alpha$ are shown.
We assume a distance modulus of 34.8. 
After subtracting the background contribution, some
magnitude bins are left with no galaxies. These are indicated
by their upper errorbars (with no mean points) only.

\vskip 10pt
\noindent
{\bf Figure 6:} The $B-K$ vs.~$B-R$ colour-colour diagram
for galaxies having $15.5 < K < 18.5$.  
Aperture magnitudes (diameter $d$ = 3.0$^{\prime\prime}$) 
are used in computing the colours.  All our data are 
from the QUIRC images, using the
$K^{\prime}$ magnitudes.  
Only objects classified by FOCAS as galaxies in all three
images are included.

Different lines represent the colour$-$colour relations 
for different types of
galaxies. 
The redshift varies along the
paths shown in the diagram, from 0 (at the bottom left
extremity of each path) to the number shown in parentheses
in the legend.  The optical $K$-corrections are from
Coleman et al.~(1980) for the giant and dIrr galaxies 
and from Trentham (1997b) for the dSph galaxies.
The $K$-band K-corrections come from Huang et al.~(1997).
The position in this colour-colour diagram of a giant
elliptical at the redshift of Coma is marked by X.
We estimate uncertainties corresponding to 0.1 and 0.15 mag.
in $B-R$ and $B-K$ colours respectively.

\vskip 10pt
\noindent
{\bf Figure 7:}       
The $B-K$ histogram of galaxies in the Coma core
field with $1.2 < B-R < 1.6$. 
These are the
galaxies which, based on their optical colours, are
most likely to be cluster members, and have
magnitudes faint enough to be in the part of the
LF that is steeply rising.  

The background contamination is small.  Only three
galaxies in the 15$^{\rm h}$ 40$^{\circ}$ background
field (this has an area of about 20\% of the 
Coma core field) satisfy the requirements for inclusion
here.  There have $B-K$ of 5.14, 5.53, and 5.99. 

\vskip 10pt
\noindent
{\bf Figure 8:}
The luminosity function of the galaxies shown
in Figure 7 with $B-K < 5$, assuming
no background contamination. 
As in Figures 6 and 7, $d$ = 3.0$^{\prime\prime}$ 
aperture magnitudes are
used; for these faint galaxies. This is a good
approximation to the total magnitudes.

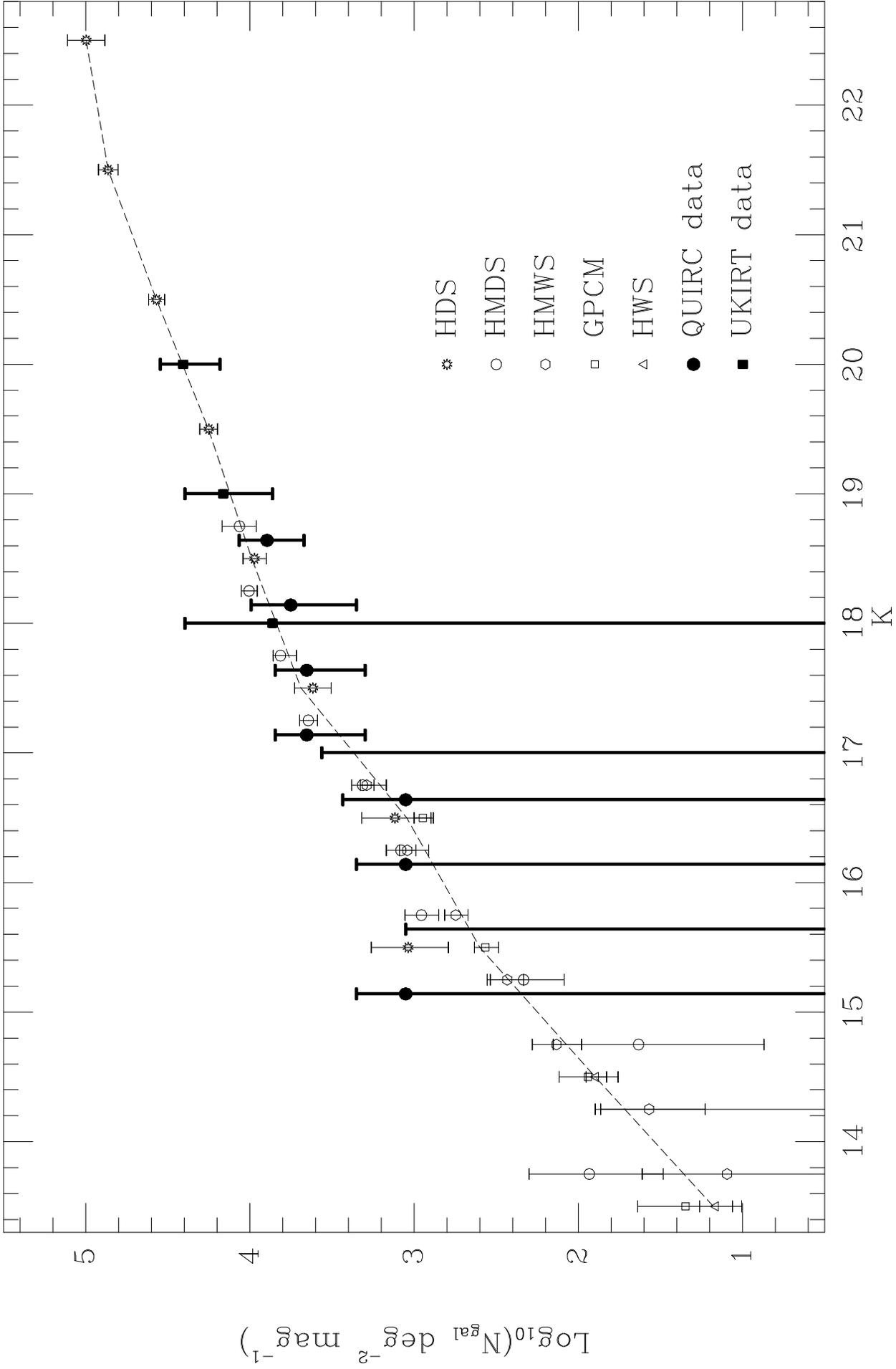
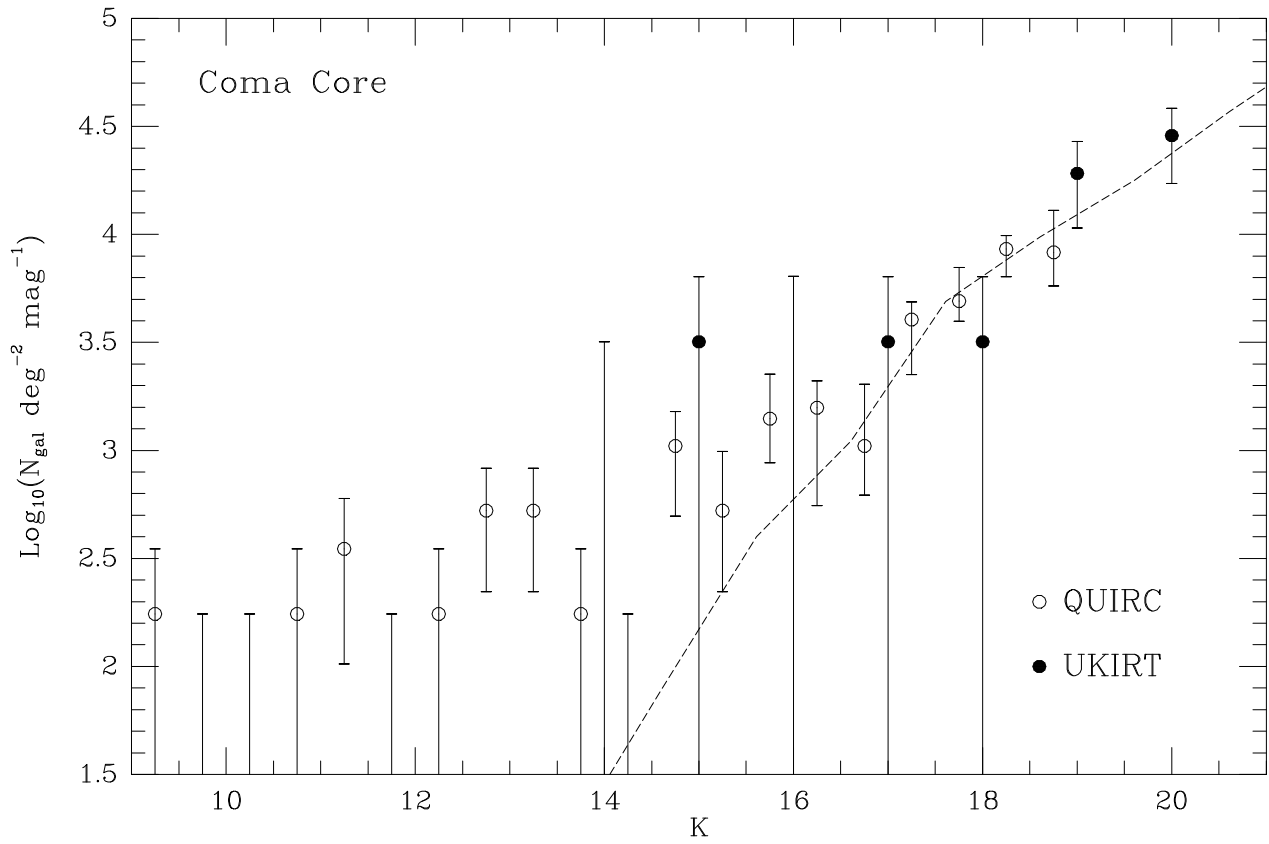
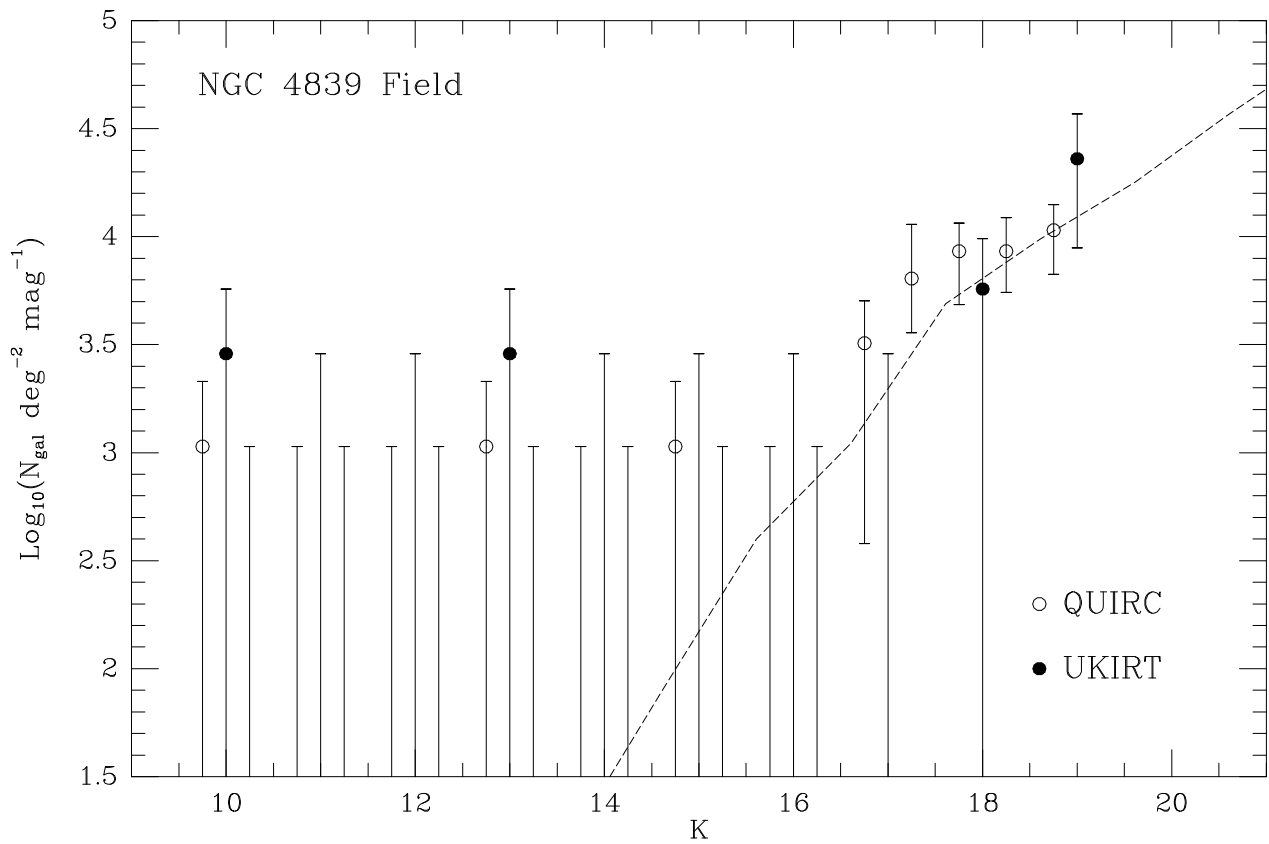
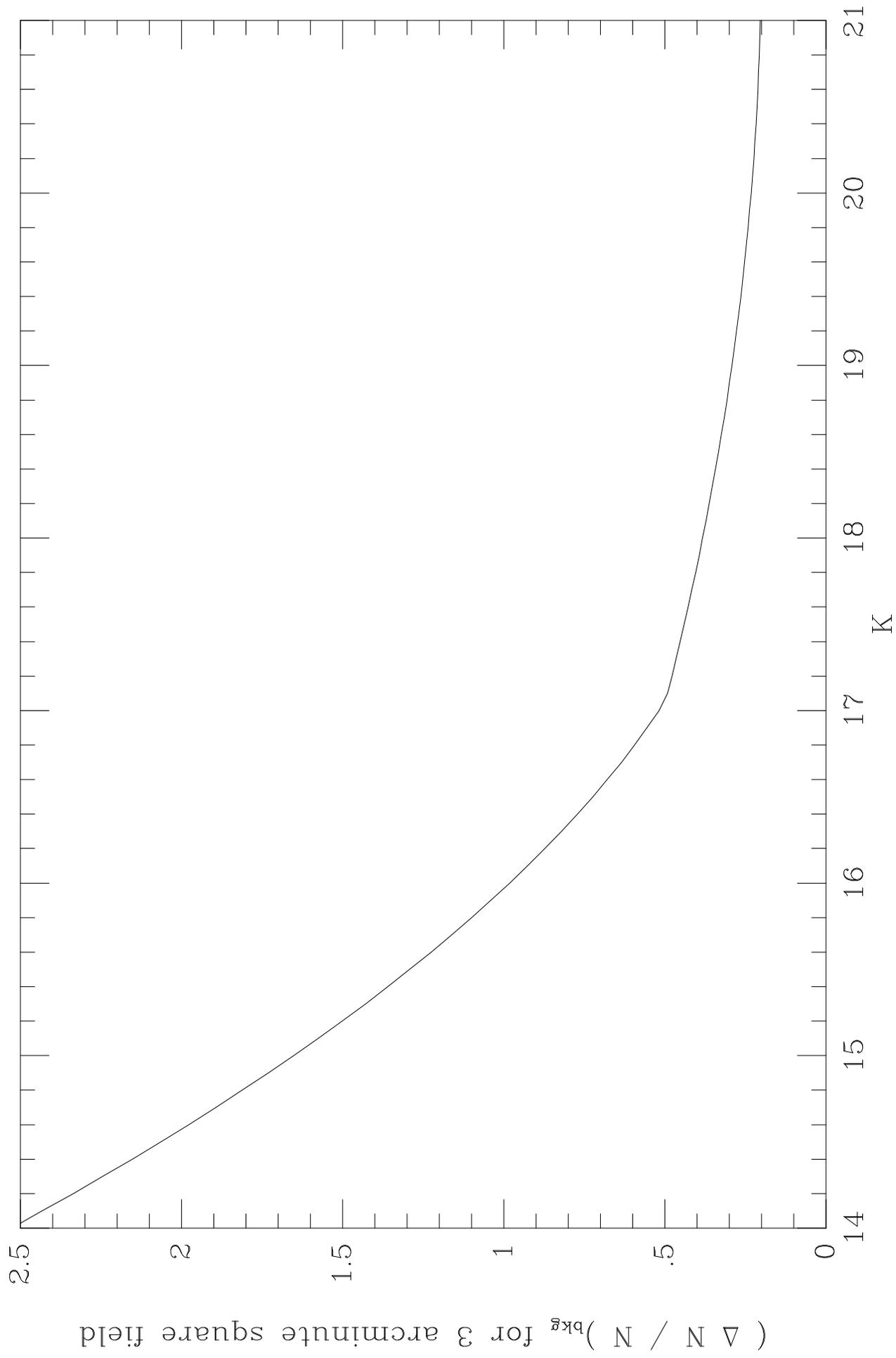
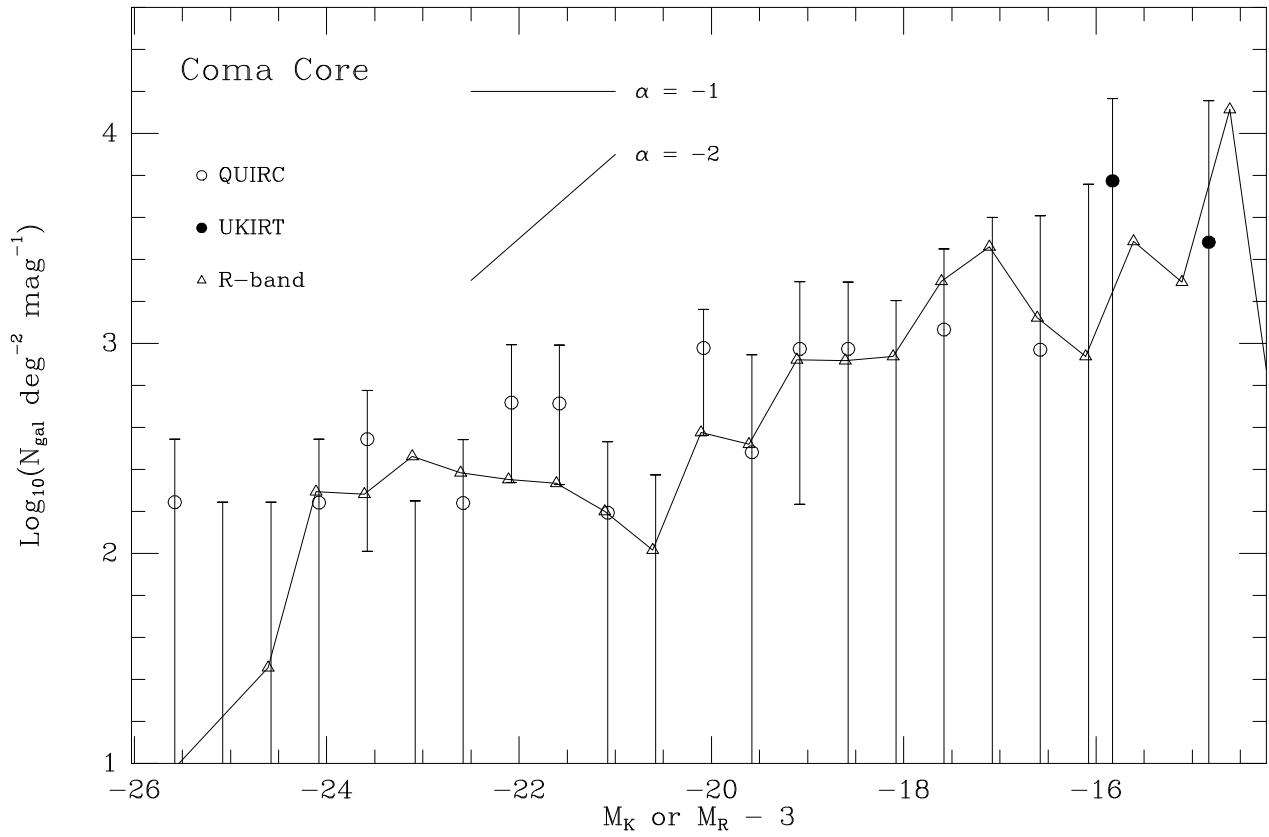
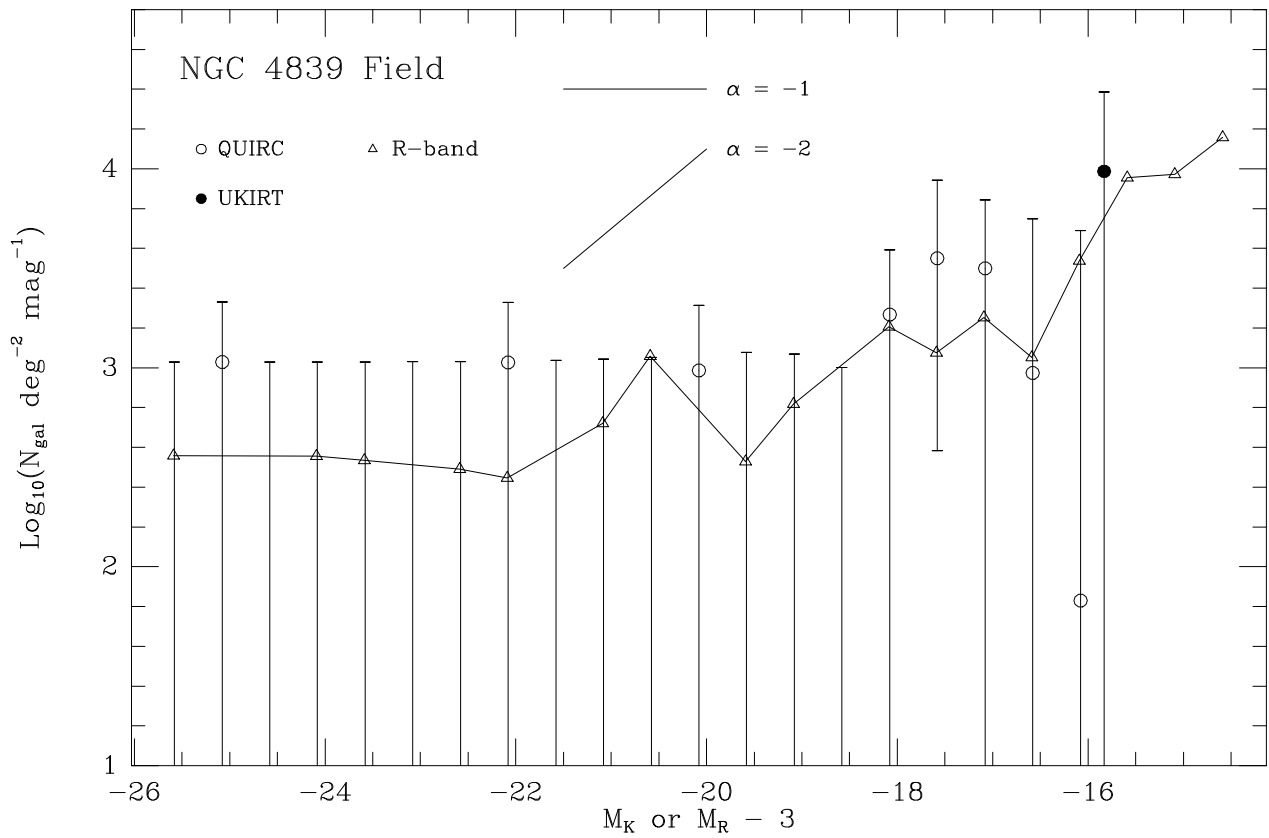
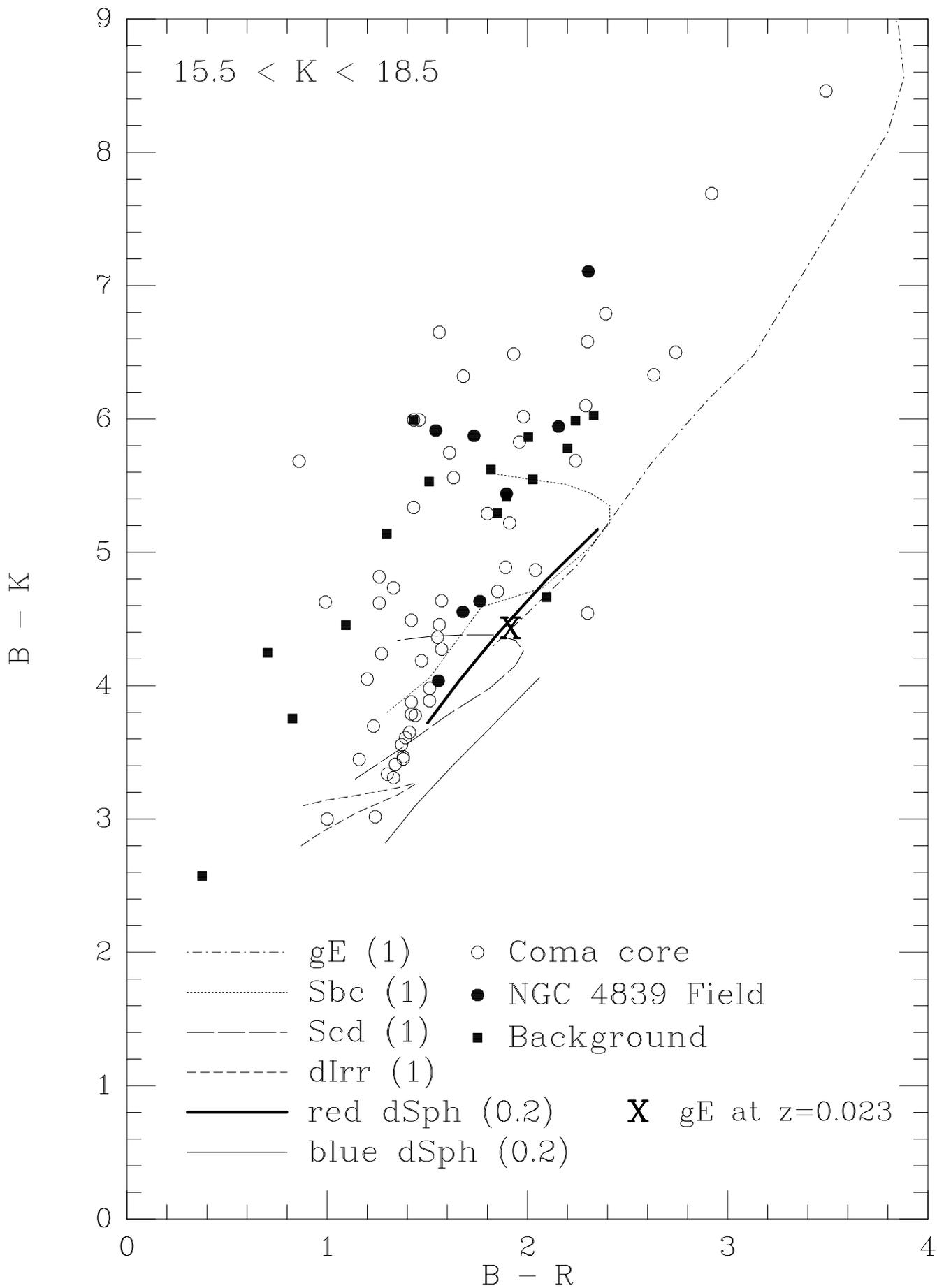
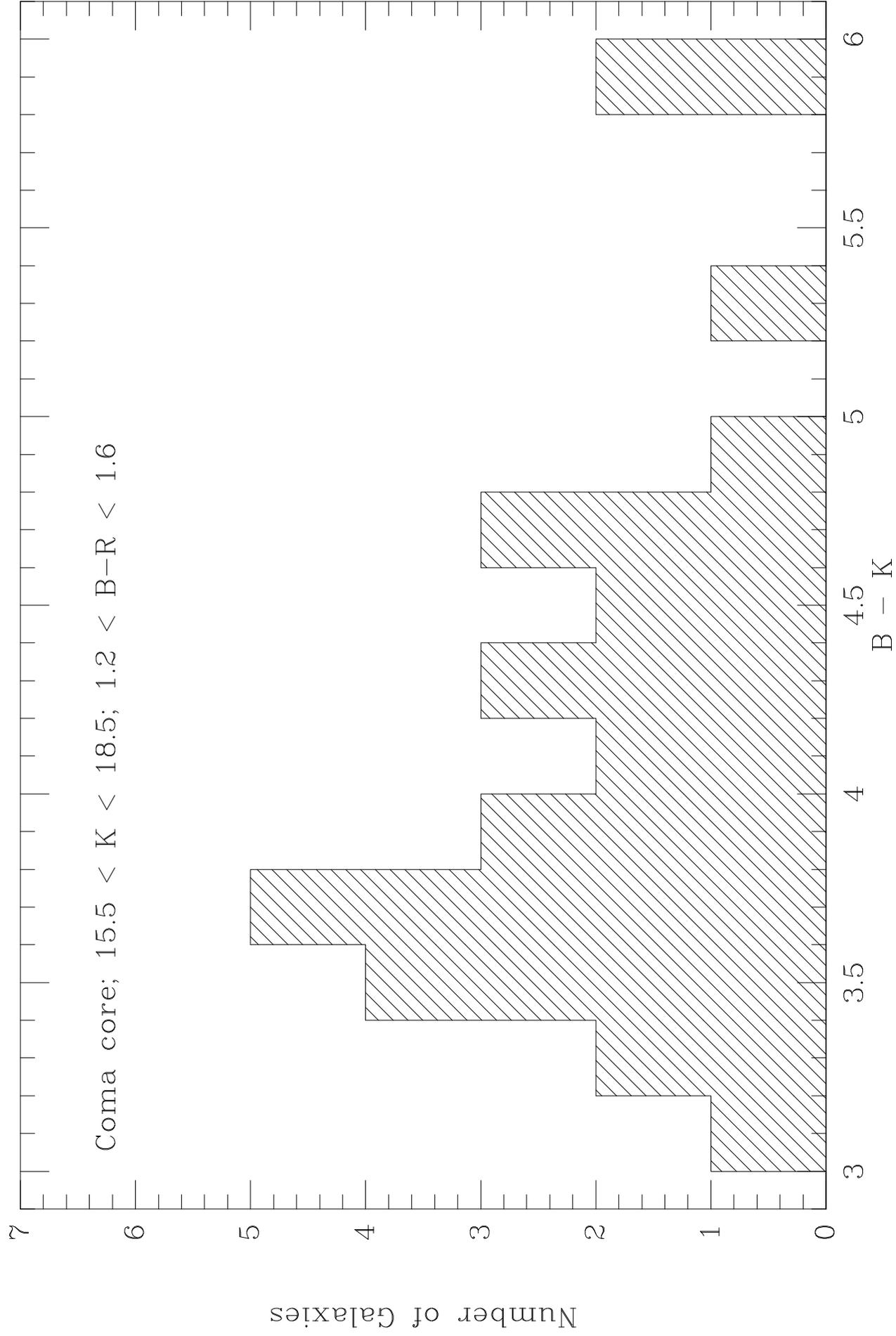
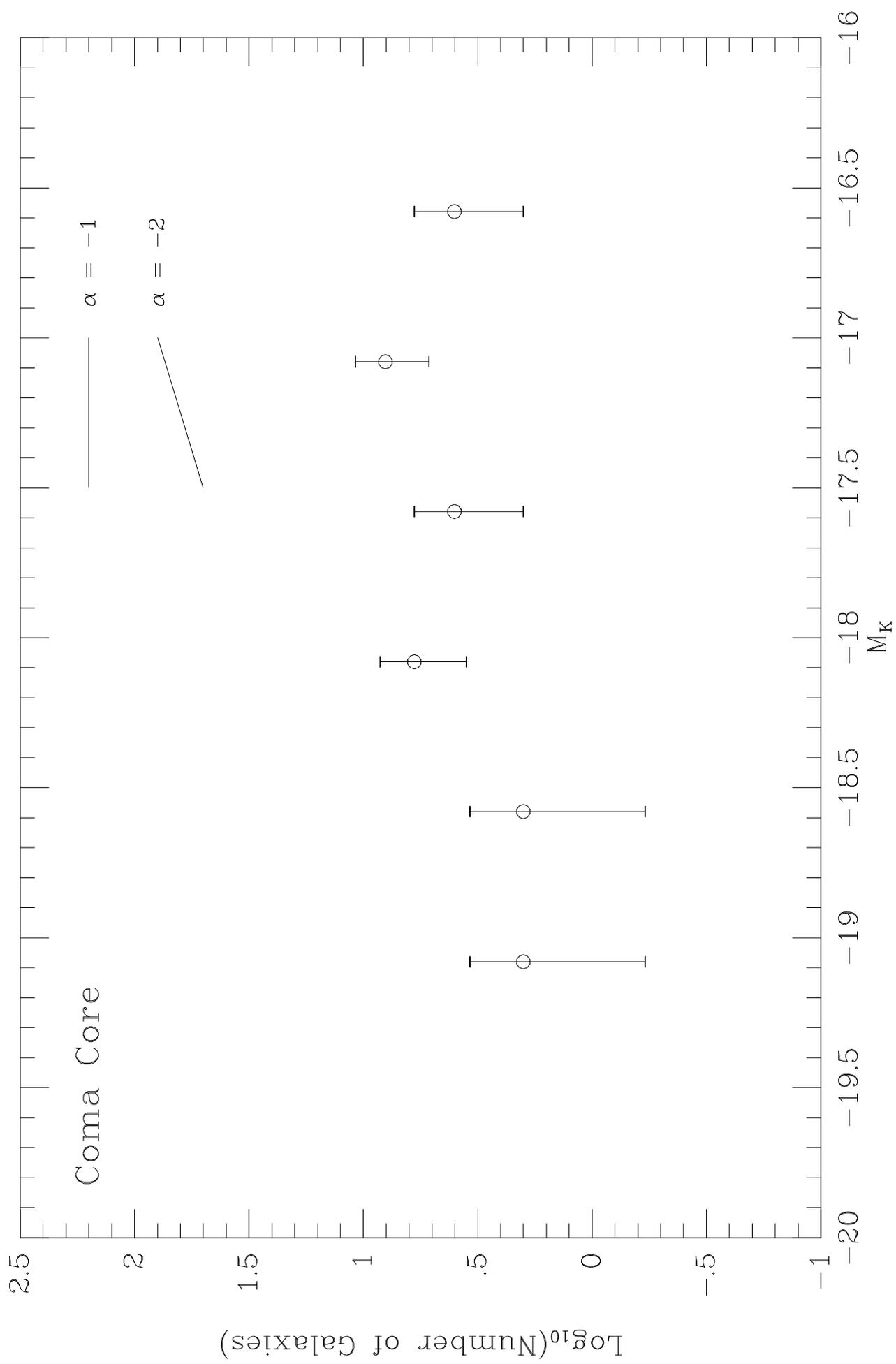
\end{document}